\documentclass[twocolumn,aps,showpacs,preprintnumbers,amsmath,amssymb]{revtex4}

\usepackage{epsf}
\usepackage{graphicx}
\usepackage{dcolumn}
\usepackage{bm}

\begin{document}
\title{Space-charge mechanism of aging in ferroelectrics:  
an exactly solvable two-dimensional model}
\author{Yuri A. Genenko} \email{yugenen@tgm.tu-darmstadt.de} %
\affiliation{Institute of Materials Science, %
Darmstadt University of Technology, %
Petersenstr. 23, 64287 Darmstadt, Germany}%

\date{\today}

\begin{abstract}

A mechanism of point defect migration triggered by local depolarization fields is shown 
to explain some still inexplicable features of aging in acceptor doped ferroelectrics. 
A drift-diffusion 
model of the coupled charged defect transport and electrostatic field relaxation within a  
two-dimensional domain configuration is treated numerically and analytically. Numerical 
results are given for the emerging internal bias field of about $1 \rm \: kV/mm$ which 
levels off at dopant concentrations well below $1 \rm \: mol \%$; the fact, long ago known 
experimentally but still not explained. For higher defect concentrations a closed solution 
of the model equations in the drift approximation as well as an explicit formula for the 
internal bias field is derived revealing the plausible time, temperature and concentration 
dependencies of aging. The results are compared to those due to the mechanism of 
orientational reordering of defect dipoles. 
\end{abstract}

\maketitle

\section{\label{sec:intro}Introduction}

The phenomenon of gradual change of physical properties with time, called aging, is 
for a long time known feature of ferroelectrics, especially when acceptor doped
\cite{plessner56aging,ikegami67mechanism,takahashi70space,thomann72stabilization,%
carl78electrical,takahashi82,arlt88internal,Lohkamper1990Gauss,%
Warren1995chargetrapping,Afanasjev2001,zhang05insitu,zhang06aging,Morozov2008}. 
Aging reveals itself in quasi-logarithmic decrease of the dielectric constant with time 
\cite{plessner56aging,thomann72stabilization}, reduction of domain wall mobility 
leading to stabilization in the aged domain structure \cite{ikegami67mechanism}, 
altered shape of polarization loops in both poled and unpoled aged samples
\cite{takahashi70space,carl78electrical,takahashi82,Afanasjev2001,zhang05insitu,%
zhang06aging,Morozov2008} and related indications. 
A characteristic aging time, $\tau$, a clamping pressure on the domain walls, $P_{cl}$, 
and an internal bias field, $E_{ib}$, were introduced as parameters quantifying aging
\cite{carl78electrical,takahashi82}.

In past three decades several 
concepts were developed \cite{TagantsevReview,dawber05physics} to explain the aging 
phenomena in terms of domain splitting \cite{ikegami67mechanism}, space charge formation 
\cite{takahashi70space,thomann72stabilization}, electronic charge trapping at domain 
boundaries \cite{Warren1995chargetrapping,Afanasjev2001}, ionic drift 
\cite{hage80_02,lamb86_02,scott87activation,Morozov2008} or reorientation of defect dipoles 
\cite{arlt88internal,Lohkamper1990Gauss,zhang05insitu,zhang06aging}. The latter concept  
based on the widely recognized mechanism of gradual orientation of defect dipoles 
formed by the charged acceptor defects and oxygen vacancies has suggested probably most 
successful quantitative explanation of many features relevant to aging and fatigue in 
ferroelectrics, particularly, plausible time and temperature dependencies of $E_{ib}$. 
Nevertheless, some long standing questions remain open, most pronounced among them the 
dependence of $\tau$ and $E_{ib}$ on the defect concentration 
\cite{carl78electrical,takahashi82}. Resulting from the individual cage motion of an 
oxygen vacancy around an acceptor defect insensitive to presence of other defect dipoles 
in the dipole reorientation model \cite{arlt88internal,Lohkamper1990Gauss} $\tau$ is 
expected to be independent on the acceptor concentration. Similarly, independent 
contributions of different defect dipoles to the clamping pressure in this model have to 
result in $E_{ib}$ directly proportional to the concentration. However, the experimentally 
observed saturation of $E_{ib}$ at medium concentrations $<1 \rm \: mol \%$ as well as 
distinct concentration dependence of $\tau$ 
\cite{carl78electrical,takahashi82,arlt88internal} provides 
indications of some collective mechanism of aging.

In this paper, therefore, we prove an alternative, space charge mechanism to explain
the above mentioned features of aging. It may also be related to the self-polarization
phenomenon and internal field-induced, migratory polarization observed in thin
ferroelectric films \cite{Afanasjev2001,Kholkin1998}. Recently, a model quantifying the
space charge mechanism was advanced \cite{lupascu06aging,Genenko-ferro2007,Genenko-ferro2008} 
which shows that the clamping pressure  $P_{cl}\simeq 1 \rm \: MPa$ 
and the field $E_{ib}\simeq 1 \rm \: kV/mm$  comparable with experiments can result from 
the formation of space charge zones near charged domain boundaries assuming small but 
finite mobility of charged defects. Aging was studied first for low defect concentrations 
about $0.01 \rm \: mol \%$ \cite{lupascu06aging,Genenko-ferro2007}, and the effect of 
anisotropy was considered \cite{Genenko-ferro2008}. Here we apply the isotropic, 
two-dimensional version of this model \cite{Genenko-ferro2007} to study aging of 
unpoled ferroelectrics for a wide range of acceptor concentrations and present 
numerical and analytical results on temperature, concentration and time dependencies 
of $E_{ib}$. Gradual change of material properties under the effect of the external dc 
or ac electric field, e.g. fatigue phenomenon, is not considered at this stage.

\section{\label{sec:generalmodel} Model of a ferroelectric grain}

Main assumptions of the model \cite{Genenko-ferro2007} used here are: 
(a) availability of mobile charged defects in amount sufficient to substantially 
compensate the bound charge at the domain boundaries and (b) presence of strong 
local depolarization fields in the unpoled ferroelectric material.

Let us consider the assumption (a). The conductivity of perovskites has been extensively 
studied during the past two decades \cite{Choi1986BTOChemistry,waser91bulk,%
Brennan1995,Raymond1996perovskitechemistry,DMSmyth2003,Guo2005conductivity,Ohly2006}.
It was established that, depending on temperature and partial pressure of oxygen, 
perovskites may exhibit ionic or electronic conductivity, so that the n-type 
conductivity prevails under reducing conditions and the p-type in oxidizing atmospheres 
\cite{Choi1986BTOChemistry,Raymond1996perovskitechemistry,DMSmyth2003}. 
The ionic conductivity prevails between the above two areas of electronic conductivity 
and is clearly dominated by oxygen vacancies. It was shown also that the bulk mobilities 
of electrons and holes are not activated in $\rm BaTiO_3$ and exceed the activated 
mobility of vacancies by many orders of the magnitude. Nevertheless, at atmospheric 
partial oxygen pressure and not very high temperatures, the concentration of electronic 
carriers remains less than the concentration of vacancies by many orders of the magnitude 
and is by far not sufficient to compensate the polarization bound charge. Moreover, the 
bulk concentration of electronic carriers is so small that the corresponding Debye 
screening length strongly exceeds the typical grain size. This allows one to entirely 
neglect electronic screening of the local bound charges. The same applies also to ambient 
electronic carriers in the intergranular space though their concentration may exceed the 
bulk one by few orders of the magnitude \cite{Guo2005conductivity}. The balance between 
the electronic and ionic species may be substantially changed locally right near the 
charged domain boundaries because of the electronic band bending by very strong local 
depolarization fields; however, this possibility depends on the reduction potentials of 
certain metal dopants~\cite{Gallardo2008}. In our study we suppose for simplicity that 
concentration of electronic carriers is negligible in and outside the ferroelectric 
grains. The oxygen vacancy concentration is usually fixed by acceptor defects even in the 
nominally undoped materials, since these are as a rule unintentionally acceptor doped 
\cite{Raymond1996perovskitechemistry,DMSmyth2003,Guo2005conductivity}. 
For these reasons, we will assume in the following that only oxygen vacancies 
participate in charge migration.

Consider now the assumption (b). In the perfectly ordered domain system 
the bound charges at the domain boundaries can be fully compensated resulting in 
complete suppression of the depolarization fields inside the bulk ferroelectric. 
Experimental studies show that it is not the case in real, disordered systems. 
Considering the high resolution images of domain patterns in $\rm BaTiO_3$ and 
lead zirconate titanate (PZT) samples by scanning transmission electron 
microscopy~\cite{Gregg2006}, by bright field 
transmission electron microscopy~\cite{SchmittTEM2007} and by secondary electron 
spectroscopy~\cite{FarooqSEM2008} one can observe numerous places where the 
uncompensated bound charges should emerge and, consequently, local depolarization 
fields can be present. Three relevant circumstances can be identified, namely, when 
a domain array a) ends up in an unpolarized area inside the grain, b) meets a domain 
wall of another domain array instead of charged domain faces, c) ends up at the grain 
boundary contacting the wide enough unpolarized integranular area. To investigate 
quantitatively the consequencies of the presence of uncompensated bound charges in the 
material the third of the above mentioned cases, (c), will be representatively studied 
in the following. Since the depolarization field of a domain array decays exponentially 
at the distance equal to the domain width $a$ \cite{lupascu06aging,Genenko-ferro2007}, 
it suffices to assume the intergrain spacing larger than $a$ to get the neighbour 
grains electrically independent. That is why, for studying the effects of local 
depolarization fields, a model of a single ferroelectric grain surrounded by 
unpolarized medium may be chosen.

\begin{figure}[htbp]
\begin{center}
    \includegraphics[scale=0.4]{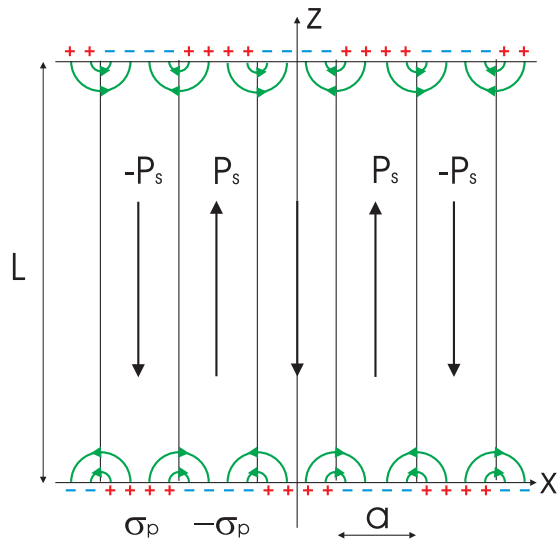}
    \caption{Scheme of a 2D-array of $180^{\circ}$-domain walls in a quadratic 
             ferroelectric grain occupying the space $|x|<L/2,\,0<z<L$.  Straight arrows 
             show the direction of the polarization and curved arrows the approximate 
             direction of the local electric field.}
\label{grain}
\end{center}
\end{figure}

Let us start with a quadratic ferroelectric grain of zero total polarization surrounded 
by an infinite dielectric medium. We imagine a two-dimensional periodic array of domains 
in the grain cut by the interfaces with dielectric, $z=0$ and $z=L$, perpendicular to the 
direction of spontaneous polarization which is along the $z$-direction of a Cartesian 
coordinate system $x, y, z$ (Fig.~\ref{grain}). 
The system is supposed to be uniform in the $y$-direction so that no variable is 
$y$-dependent. If the length of the domains $L$ along the $z$-axis is much larger than 
their width $a$ along the $x$-axis, which is typically the case in experiments, electric 
field lines are effectively closed at the same side of the grain. Finite-element 
simulations of the electric field in the finite domain array shows a virtually periodic 
field pattern along the $x$ axis everywhere but the very ends of the array 
\cite{lupascu06aging,Genenko-ferro2008}. That is why we can consider a periodic domain 
array infinite along the $x$ axis as a representative model for a finite multi-domain 
grain. This model configuration is well-known in the physics of polarized media and was 
used for the study of equilibrium and dynamic properties of ferromagnetic
\cite{Kittel1946,LandauElectrodynamicsContinuum} and ferroelectric
\cite{Mitsui1953} materials.

Furthermore, since both components of the depolarization field exponentially decrease 
towards the interior of the grain along the $z$-axis 
\cite{lupascu06aging,Genenko-ferro2007}, transport of the charged defects driven by the 
field is expected to occur in the vicinity of the grain boundaries $z=0$ and $z=L$. 
Considering charge migration near the interface $z=0$ we can therefore assume the domains
to be infinite along the $z$-axis without introducing a substantial error. Apparently, 
the same process of charge separation occurs at the other grain boundary, $z=L$, too. 
Calculating the forces exerted upon the domain walls, both ends of the domains 
must be taken into account.

Hence, to study the depolarization field induced charge migration it is sufficient to
consider the interface $z=0$ between the domain array occupying the ferroelectric half 
space $z>0$ and the dielectric medium occupying the half space $z<0$. For simplicity, 
both media are assumed to be isotropic and characterized by the relative dielectric 
permittivities $\varepsilon_f$ and $\varepsilon_d$, respectively. Due to polarization, 
the domain faces at $z=0$ are alternatively charged with the bound surface charge density 
\cite{LandauElectrodynamicsContinuum}
\begin{eqnarray}
\label{face-charge}
\rho_b(x,z)=\sigma_p\delta(z)\sum_{n=-\infty}^{\infty}(-1)^n\nonumber\\
\times\theta\left(\frac{a}{2}-an+x\right)
\theta\left(\frac{a}{2}+an-x\right),
\end{eqnarray}
\noindent where $\sigma_p = |\mathbf{P}_s|$ is the spontaneous polarization value,
$\delta (z)$ and $\theta  (x)$ are the Dirac $\delta $-function and the Heaviside unit 
step function, respectively. The choice of the origin in the middle of the positively 
charged domain face makes the problem also bilaterally symmetrical.

Migration of charge carriers is governed by the drift-diffusion equation which is 
often used for description of electronic transport in semiconductors \cite{Sze}:
\begin{equation}
\label{continuity}%
\partial_t c=-\nabla(\mu c {\bf E}) + D\triangle c \,, 
\end{equation}
\noindent where ${\bf E}(x,z,t)$ is the local electric field at the time $t$, $c(x,z,t)$ 
is the concentration of positively charged, mobile species, $\mu$ and $D$ are 
their mobility and diffusivity, respectively. We assume, for simplicity, that the latter 
two quantities are isotropic and connected by the Einstein relation, $D=\mu kT/q_f$ with 
$k$ the Boltzmann constant, $T$ the absolute temperature and $q_f$ the charge of the 
carriers.

The electric field ${\bf E}(x,z,t)$ is determined self-consistently by the charged faces 
of the domains and the imbalanced charge density in the bulk $\rho(x,z,t)=q_f
\left[c(x,z,t)-c_0\right]$ through Gauss' law
\begin{equation}
\label{Gauss}%
\nabla{\bf E}=\frac{q_f}{
\varepsilon_f\varepsilon_0}(c-c_0) \,,
\end{equation}
\noindent where $c_0$ is the background concentration of the immobile acceptor defects
warranting total electroneutrality, and $\varepsilon_0$ is the permittivity of vacuum.

For the boundary conditions to the system of equations (\ref{continuity}) and 
(\ref{Gauss}) we assume usual boundary conditions for the electric field at the
interface between the two media \cite{LandauElectrodynamicsContinuum} and chemical 
isolation of the grain determined by the requirement of vanishing particle current
at $z=0$:
\begin{equation}
\label{boundary-chem}%
\mu c E_z - D\partial_z c=0\,.
\end{equation}
Note that the latter condition is not crucial for the present model. Transport of 
charged species in the dielectric medium (intergranular space) and also through the 
grain boundary may be additionally included. Here, for simplicity, we consider only
redistribution of mobile defects inside the grain.

In the initial state, the system is locally neutral assuming $c(x,z,0)\equiv c_0$, 
while the electric field ${\bf E}(x,z,0)\equiv{\bf E}^0(\sigma_p|x,z)$ is determined
at the beginning  solely by the surface charge density (\ref{face-charge})  and 
reads, as was calculated in Ref.~\cite{Genenko-ferro2007}, 
\begin{eqnarray}
\label{E0}
E^0_x=\frac{\sigma_p}{\pi\varepsilon_0 
(\varepsilon_f + \varepsilon_d )}
\ln{\left[\frac{\cosh(\pi z/a)+\sin(\pi x/a)}
{\cosh(\pi z/a)-\sin(\pi x/a)}\right]} ,\nonumber\\
E^0_z=\frac{2\sigma_p}{\pi\varepsilon_0}
\frac{1}{(\varepsilon_f + \varepsilon_d )}
\arctan{\left[\frac{\cos(\pi x/a)}
{\sinh(\pi z/a)}\right]}.
\end{eqnarray}
\noindent The latter equations together with boundary conditions and 
Eqs.~(\ref{continuity}) and (\ref{Gauss}) complete the formulation of the problem 
which can now be treated numerically.

\section{\label{sec:numsolution}Numerical evaluation of the internal bias field}

Numerical solution of the problem is performed using a simple direct Euler scheme  
described in Ref.~\cite{Genenko-ferro2007}. Due to the periodicity and the bilateral 
symmetry of the task, it is sufficient to consider the charge redistribution within 
the area $0<x<a$ contaning a single domain wall at the plane $x=a/2$. Space and time are 
discretized. At every time step, the change in the carrier concentration is computed   
from the previous values of the concentration and the electric field using 
Eq.~(\ref{continuity}). Updated values of the field are then calculated  using 
Eq.~(\ref{Gauss}). The evaluation is repeated until convergence is achieved. 

As an example, we consider now the aging process in unpoled $\rm BaTiO_3$ doped with a
bivalent acceptor, e.g. Ni, Ca or Mg. For numerical simulations, the material 
parameters at temperature $45^{\circ} \rm \:C$  are taken from 
Refs.~\cite{arlt88internal,jaff71} with the aim to compare results 
directly with those of the theory of dipole reorientation. Namely, it is assumed that 
$P_s=2.7\cdot 10^{-1} \rm \: C/m^2$, $\varepsilon_f = 200$,  $a=0.5\rm \: \mu m$. 
Positively charged oxygen 
vacancies are assumed as mobile species with $q_f$ twice the elementary charge and the 
mobility $\mu=8.4\cdot 10^{-22} \rm \: m^2/Vs$ implying activation energy of  
$E_a=1.1\rm \: eV$~\cite{waser91bulk,Raymond1996perovskitechemistry,DMSmyth2003,Ohly2006}.

Note that, for the dielectric permittivity the intrinsic, lattice value is 
taken~\cite{jaff71} since the charge migration occurs at the mesoscopic scale of the 
domain width. The directly measurable macroscopic permittivity may achieve much higher 
values due to the large contribution of the domain walls~\cite{WangGiant2007} which 
does not apply in the considered problem. 
Account of anisotropy of crystalline $\rm BaTiO_3$ would result in enhancement of the 
field magnitude by the factor $\sqrt{\varepsilon_a/\varepsilon_c}$ and in reduction
of the field penetration depth by the same 
factor~\cite{Genenko-ferro2007,Genenko-ferro2008}, where $\varepsilon_a=2180$ and 
$\varepsilon_c=56$  are the principal values of the permittivity tensor~\cite{Zgonik}. 
Since the electrostatic energy is quadratic in the field this would entail the increase 
in the values of the clamping pressure and the internal bias field by the factor 
$\sqrt{\varepsilon_a/\varepsilon_c} \simeq 6$. However, for 
simplicity and comparability with the results of the dipole reorientation 
model~\cite{arlt88internal,Lohkamper1990Gauss}, we assume here the above introduced 
isotropic permittivity $\varepsilon_f$. For the dielectric medium outside the 
ferroelectric grain we take the same but non-polarized material with 
$\varepsilon_d=\varepsilon_f$.

The system reveals two typical time scales: the drift time 
$\tau_{\mu}=a/\mu E^0_z \simeq 7.8\cdot 10^6\rm \:s$ and the diffusion time 
$\tau_D=a^2/D\simeq 2.2\cdot 10^{10}\rm \:s$ so that the ratio 
$\beta = \tau_{\mu}/\tau_D\simeq 3.6\cdot 10^{-4}$ characterizes the contribution of
diffusion to Eq.~(\ref{continuity})~\cite{Genenko-ferro2007}. Though very small, this 
contribution cannot be neglected because it influences the structure of space charge 
zones and provides compatibility with the boundary condition (\ref{boundary-chem}). 
However, because of very small $\beta$, large gradients of concentration arise near 
the negatively charged face of the domain at $z=0,\,a/2<x<a$ which make the numerical 
procedure unstable. Since the parameter $\beta$ affects nothing but the thickness of 
the positively charged layer piled up near the negatively charged domain 
face~\cite{Genenko-ferro2007} we assume in computations $\beta=5\cdot 10^{-2}$ keeping 
in mind that this may lead to mistakes if the thickness of the space charge zone in 
front of the positively charged domain face becomes comparable with $\beta a$.  
Details on the field and concentration profiles and their evolution with time are 
exemplary presented in Ref.~\cite{Genenko-ferro2007} for low dopant concentrations 
about $c_0=0.01 \rm \: mol \%$. Here computations are extended over the region
from $c_0=0.01 \rm \: mol \%$ to $1 \rm \: mol \%$.

Having the charge density and the electric field calculated, the time dependent forces 
exerted upon domain walls can be evaluated. The loss of domain wall mobility, 
characteristic of aging, results from relaxation of the energy of the electrostatic 
depolarization field due to piling up of the charged defects at the charged domain
faces. Distribution of the energy of this field and of the consequent clamping pressure 
along the domain wall is very nonuniform, peaking at the domain 
boundaries~\cite{Genenko-ferro2007}. The average clamping pressure preventing the 
displacement of the domain wall from the energy minimum and the corresponding internal
bias field may be estimated as follows.

The thermodynamic force exerted upon the domain wall can be defined as the derivative 
of the energy of the electrostatic field on the domain wall diplacement. This derivative
may be roughly estimated from the difference between the energy of the initial state of 
the system displayed in Fig.~\ref{grain} and the fully polarized state achieved by the 
virtual displacement of domain walls over the distance of $a/2$.

The energy of the electrostatic field per one half of the domain length reads
\begin{equation}
\label{field_energy}%
W[{\bf E}]=\frac{1}{2}\varepsilon_0 \varepsilon_f \int_{-\infty}^{L/2} dz \int_0^a dx\,
 {\bf E}^2 ,
\end{equation}
\noindent while the other half of the domain at $z>L/2$ contributes the same amount of
the energy for symmetry reasons. In the virgin state of the grain, the depolarization 
field is represented by Eqs. (\ref{E0}), and the energy of this field per one period of 
the structure is given by the well known formula 
\cite{Kittel1946,LandauElectrodynamicsContinuum,Mitsui1953}
\begin{equation}
\label{virgin_energy}%
W[{\bf E}^0(\sigma_p)]= 0.85 \frac{\sigma_p^2 a^2}{4\pi \varepsilon_0 \varepsilon_f}.
\end{equation}
In the course of aging, the electric field transforms to 
${\bf E}={\bf E}^0(\sigma_p)+\Delta {\bf E}$, where $\Delta {\bf E}$ is the contribution 
to the field due to the charge carrier migration. The aged state of the structure in
Fig.~\ref{grain} serves as the initial state with the energy $W[{\bf E}]$ in the 
clamping force calculation.

Consider now the virtual pairwise displacement of the domain walls to each other until 
they meet which leads to the full polarization of the system in the positive $z$ 
direction. The resulting uniform bound charge at the grain boundaries at $z=0$ and $z=L$ 
generates then the uniform depolarization field 
${\bf E}_d = (0,0,-P_s/\varepsilon_0 \varepsilon_f),\,0<z<L$, so that the total electric 
field becomes equal to ${\bf E}_f = {\bf E}_d +\Delta {\bf E}$. 
The energy of the electrostatic field in this state amounts to $W[{\bf E}_f]$. The 
clamping pressure on the domain walls related to aging is provided by the time-dependent
part of the energy difference $W[{\bf E}_f]$-$W[{\bf E}]$, namely
\begin{equation}
\label{clamping-energy}%
\Delta W_{cl}=\varepsilon_0 \varepsilon_f \int_{-\infty}^{L/2} dz \int_0^a dx
\left( {\bf E}_d -{\bf E}^0 \right)\Delta {\bf E} .
\end{equation}
\noindent Calculating here the energy gain near the domain boundary at $z=0$ the 
integration limit $L/2$ can be extended to infinity because of the exponentially fast 
descrease of the fields ${\bf E}^0$ and $\Delta {\bf E}$ in both directions of $z$ axis.

Finally, the internal bias field in the $z$ direction can be evaluated comparing the 
force $2P_s {\mathcal E} L$~\cite{Nechaev} exerted upon the domain wall by an external 
field ${\mathcal E}$ 
with the clamping force $2\Delta W_{cl}/(a/2)$ accounting now for 
both domain boundaries. This results in the estimation
\begin{equation}
\label{Eib-formula}%
E_{ib} \simeq \frac{2}{aLP_s} \Delta W_{cl}.
\end{equation}

\begin{figure}[htbp]
\begin{center}
    \includegraphics[scale=0.3]{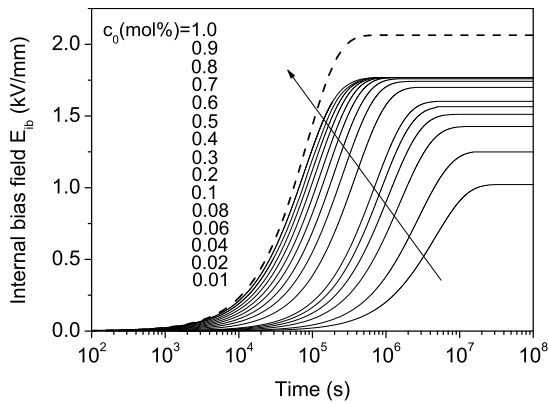}
    \caption{$E_{ib}$ as a function of time for variable acceptor concentration $c_0$ 
which increases monotonically from the lower to the upper curve (solid lines). The 
dashed line presents the theoretical limit of high $c_0$ (see text).}
\label{CoerciveField}
\end{center}
\end{figure}

\begin{figure}[htbp]
\begin{center}
    \includegraphics[scale=0.3]{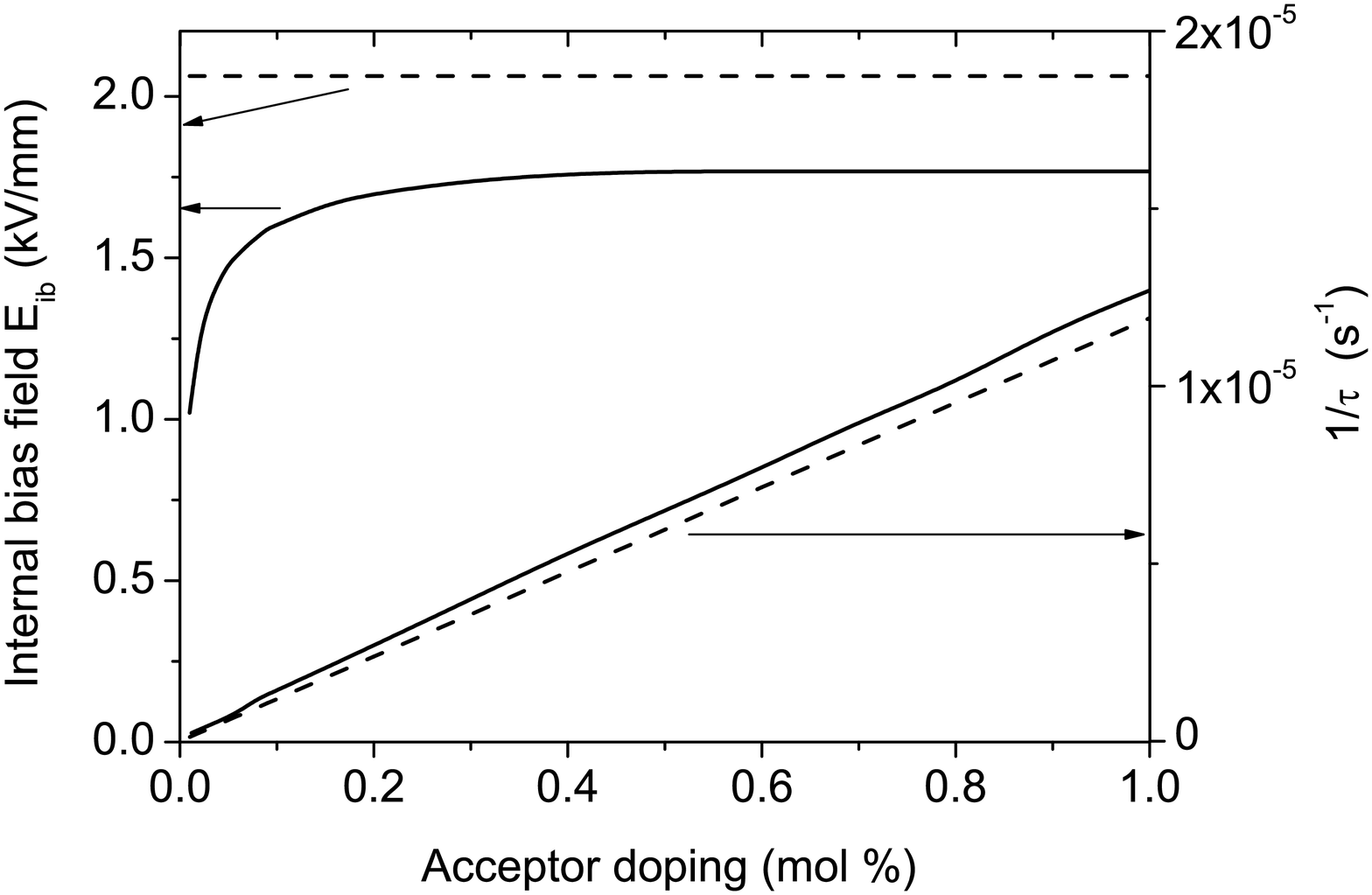}
    \caption{Saturated value of $E_{ib}$ and inverse aging time, $\tau^{-1}$, 
    determined from the inflection point of plots in Fig.~\ref{CoerciveField}, 
    are presented as functions of acceptor concentration $c_0$ by solid lines. 
    Respective theoretical values derived in the limit of high $c_0$ (see text) 
    are shown by dashed lines.}
\label{Eib-c0}
\end{center}
\end{figure}

Evaluation of the time-dependent field $E_{ib}$ assuming the typical length of the 
domain wall $L=20a$ is shown in Fig.~\ref{CoerciveField} for different dopant 
concentrations.    
All the curves demonstrate saturation of $E_{ib}$ after a some characteristic (aging) 
time $\tau$ which decreases with the growing concentration. The final, equilibrium value
of $E_{ib}$ first increases with the concentration but then levels off well below 
$c_0\simeq 1.0\rm \: mol \%$ as is seen in Fig.~\ref{Eib-c0}.  
The inversed aging time $\tau^{-1}$ rises almost linearly with the increasing 
doping in the whole range of concentrations studied. Saturating dependence of $E_{ib}$ 
on $c_0$ is clearly seen in experiments on the acceptor doped PZT 
\cite{carl78electrical,takahashi82}; indications of both above described $E_{ib}$ 
and $\tau$ dependencies are observable on the acceptor doped barium titanate 
too~\cite{arlt88internal}.

Saturation of $E_{ib}$ as well as decreasing of aging time with growing concentration 
$c_0$ has a simple physical reason. The depth $\Delta$ of the space charge area emerging 
near the positively charged domain boundary is defined by the amount of carriers needed 
to fully compensate the surface bound charge and is of the order of $\sigma_p /(q_f c_0)$. 
For the concentration of $c_0 = 1 \rm \: mol\%$ it is about $5.4 \rm \: nm$ which is 
two orders of the magnitude smaller than the domain width $a$. This means that, at so 
high concentrations, the electrostatic depolarization field is compensated virtually 
in the whole specimen volume due to the charge migration so that the maximum possible 
energy gain and, consequently, the maximum magnitude of $E_{ib}$ is achieved which is
concentration independent. At smaller concentrations, especially below 
$c_0 = 0.1 \rm \: mol\%$, the extended space charge zone exists near the positively 
charged domain face where the depolarization field is still 
present~\cite{Genenko-ferro2007}. This makes the field suppression incomplete, so that 
the energy gain and, consequently, $E_{ib}$ is not maximum and increases with the 
concentration. The characteristic time of aging is determined, in turn, by the distance 
$\Delta$ the charge carriers have to cover and is obviously proportional to 
$\Delta/\mu E^0_z \sim 1/c_0$. The delineated dependencies of both $E_{ib}$ and $\tau$ 
are clearly seen in Figs. \ref{CoerciveField} and \ref{Eib-c0}.

\section{\label{sec:ansolution}Analytical solution for medium and high defect 
concentrations}

If the vacancy concentration is high enough in the sense of charge compensation mechanism
discussed in the previous section, the problem may be substantially simplified by 
neglecting the diffusion contribution relevant only within the thin space charge zone. 
Then Eq. (\ref{continuity}) considered in the drift approximation reads    
\begin{equation}
\label{drift}%
\partial_t c=-\nabla(\mu c {\bf E}).
\end{equation}
\noindent

The boundary condition (\ref{boundary-chem}), where drift was outweighed by diffusion, 
does not apply anymore. To keep charge balance this condition is substituted by the 
requirement on the surface charge density $\sigma(x,t)$ which now embraces the space 
charge zones and can be obtained by integration of Eq.~(\ref{drift}) over an 
infinitesimal region near the boundary:  
\begin{equation}
\label{sigma_t}%
\partial_t \sigma(x,t)=- q_f\mu c(x,+0,t)E_z(x,+0,t).
\end{equation}
This assumption implies that the defect concentration remains constant in the bulk 
resulting in an ansatz
\begin{equation}
\label{c_ansatz}%
c(x,z,t)=c_0+\delta(z)  \sigma (x,t).
\end{equation}

Since $E_z(x,+0,t)=\sigma (x,t)/(\varepsilon_f + \varepsilon_d )$,
Eq.(\ref{sigma_t}) leads to an equation for the surface charge density 
\begin{equation}
\label{sigma}%
\partial_t \sigma(x,t)=-\sigma(x,t)/\tau_r 
\end{equation}
with the Maxwell relaxation time 
$\tau_r =\varepsilon_0 (\varepsilon_f + \varepsilon_d )/\kappa$ where 
$\kappa=q_f\mu c_0$ is the ionic conductivity of the bulk material. An apparent initial 
condition for $\sigma(x,t)$ reads $\sigma(x,0)=\sigma_p\, \text{sign} (a/2-x)$ which 
provides a solution $\sigma(x,t)=\sigma_s (t)\, \text{sign} (a/2-x) $ with
$\sigma_s(t)=\sigma_p \exp{(-t/\tau_r )}$. The asymptotic condition 
$\sigma_s(t\rightarrow \infty )=0$ means full compensation of the bound charge.

With the ansatz (\ref{c_ansatz}), Eq.~(\ref{Gauss}) becomes homogeneous and is satisfied 
by the function ${\bf E}^0(\sigma_s(t)|x,z)$. The drift equation (\ref{drift}) is then 
self-evident satisfied in the bulk. The part of the field due to charge migration
equals consequently $\Delta {\bf E}=-{\bf E}^0(\sigma_p) [ 1 - \exp{(-t/\tau_r) }]$.

Substituting $\Delta {\bf E}$ into Eq.~(\ref{clamping-energy}) one can note
that, for symmetry reasons, the first term in the brackets does not contribute to the 
energy . Finally, an explicit formula for the internal bias field follows from Eq. 
(\ref{Eib-formula}):
\begin{equation}
\label{Eib-exact}%
E_{ib}(t)\simeq \frac{0.85}{\pi} \frac{a}{L} \frac{P_s}{\varepsilon_0\varepsilon_f}
\left[ 1 - \exp{(-t/\tau_r) } \right]. 
\end{equation}
\noindent Note that, neglecting the thickness of the space charge zone, the saturated 
(asymptotic) magnitude of $E_{ib}$ achieves the concentration independent maximum value 
determined by the electrostatic energy of the stripe domain structure 
(\ref{virgin_energy}). The aging time $\tau$ is represented by the Maxwell relaxation 
time $\tau_r$ which is, as expected, proportional to $1/c_0$. 

The dependence (\ref{Eib-exact}) is shown exemplary for the concentration 
$c_0 = 1 \rm \: mol\%$ on the Fig.~\ref{CoerciveField} (dashed line). It describes well 
an increase of the corresponding computed curve for $E_{ib}$ below the aging time but 
levels off at the magnitude approximately $15\%$ larger than the computed value. We 
interpret this difference as a result of the growing numerical mistake at high 
concentrations $c_0$ when the thickness of the space charge zone becomes of the order of 
$\beta a$ as discussed in Sec. \ref{sec:numsolution}. The Maxwell relaxation time 
(dashed line) describes well the aging time at all concentrations considered as is shown 
in Fig.~\ref{Eib-c0}.

\begin{figure}[tbp]
\begin{center}
    \includegraphics[scale=0.3]{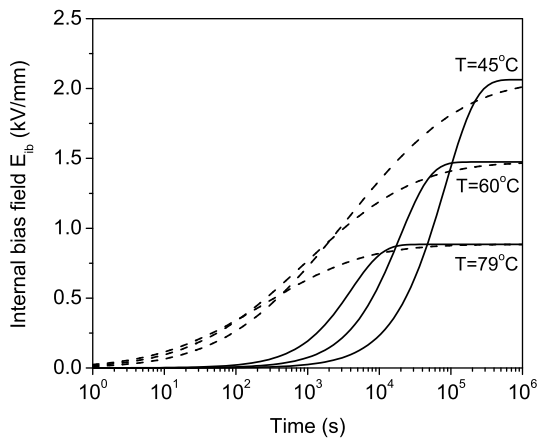}
    \caption{$E_{ib}$ as a function of time for activation energy  
             $E_a = 1.1\rm \: eV$, acceptor concentration $c_0 = 1 \rm \: mol \%$ and 
             variable temperature as indicated in the plot (solid lines). $E_{ib}$ 
             averaged over the Gauss distribution of the random energies $E_a$ 
             (dashed lines).}
\label{Eib-T}
\end{center}
\end{figure}

The field $E_{ib}$ is temperature dependent through the parameters $\varepsilon_f$ 
and $\tau_r$ of the formula (\ref{Eib-exact}). $\varepsilon_f$ changes strongly when 
temperature increases towards the ferroelectric transition that is regarded as in 
Ref.~\cite{arlt88internal}. The dependence of $\tau_r$ is a more subtle question since, 
in the considered temperature range, the directly measured total conductivity can be 
dominated by electron holes~\cite{DMSmyth2003,Guo2005conductivity,Ohly2006} because of 
their mobility much higher than the ionic one. However, the density of electronic 
carriers remains, as was discussed in Sec. III, much lower than that of oxygen vacancies 
and cannot contribute much to the bound charge compensation. That is why in evaluation 
of $\tau_r$ only the ionic conductivity $\kappa(T)=\kappa_0 \exp{(-E_a/kT)}$ is included 
with the activation energy $E_a=1.1\rm \: eV$ and 
$\kappa_0=112.8 \rm \: (\Omega\cdot cm)^{-1}$ for 
$c_0 = 1 \rm \: mol \%$~\cite{DMSmyth2003}. For these $c_0$ and $E_a$, time 
dependences of $E_{ib}$ by different temperatures are presented in Fig.~\ref{Eib-T}
by solid lines. In fact, the activation energy for oxygen vacancies is not known 
accurately and usually estimated as 
$E_a= (1\pm 0.1)\rm \: eV$~\cite{waser91bulk,Ohly2006}. Assuming, as in 
Ref.~\cite{Lohkamper1990Gauss}, that $E_a$ is random variable distributed with a 
Gaussian $\sim \exp[(E_a-\bar E)^2/2s^2]$ using $\bar E=1\rm \: eV$ and  
$s/\bar E= 0.1$, $E_{ib}$ can be averaged over this 
distribution~\cite{Lohkamper1990Gauss,Tagantsev2002stretche}
resulting in a quasilogarithmic time dependencies shown by the dashed lines for 
different temperatures in Fig.~\ref{Eib-T}.

Theoretical curves in Fig.~\ref{Eib-T} represent satisfactorily experimental time and 
temperature dependencies of $E_{ib}$ \cite{arlt88internal,Lohkamper1990Gauss}. Somewhat 
overestimated value of $E_{ib}$ could be adjusted by the factor $a/L$ which is, in fact, 
the only fitting parameter in this theory. In the performed calculations it was taken 
equal to $0.05$ but it is, in fact, slightly dependent on the grain size and lies 
between $0.02$ and $0.05$ \cite{HoffmannActa2001}. The parameter $a/L$ may be also used
to account for the fact that not every domain array is pinned by the local charges. 
The domain arrays which perfectly match other domain arrays, compensating bound charges,
are stiffly coupled to each other. This means that for $L$ some effective domain length 
$L_{eff}$ can be taken which may exceed $L$ few times but is hardly larger than 
the grain size.

The curves in Fig.~\ref{Eib-T} resemble those in the theory of dipole reorientation 
\cite{arlt88internal,Lohkamper1990Gauss} though the mechanisms involved are completely 
different which reveals itself in different dependencies on the doping concentration. 
Note that the described here space charge mechanism of domain pattern fixation near 
domain boundaries does not preclude at all the dipole reorientation mechanism which can 
still be valid in the bulk of the grain.

\section{\label{sec:conclusions}Discussion and Conclusions}

The two-dimensional model of depolarization field induced charge migration has been 
presented which explains plausibly the time, temperature and doping dependencies 
of the internal bias field in aging ferroelecrics. Saturation of this field as well 
as descrease of the aging time with the increasing dopant concentration is in
agreement with experimental observations  
\cite{carl78electrical,takahashi82,arlt88internal}. This is in contrast to the theory 
of defect dipole reorientation
\cite{arlt88internal,Lohkamper1990Gauss} based on the picture of individual cage 
motion of oxygen vacancies which predicts the internal bias field proportional to
and the aging time independent on the dopant concentration.

A reasonable question arises: whether account of interaction between defect dipoles 
could modify the dipole rotation theory so that it explains the concentration dependence 
properly. This possibility should indeed be carefully studied. It is known, at least, 
that, at high vacancy concentrations, substantial structural changes in the subsystem 
of vacancies may occur \cite{Steinsvik1997}. This happens, however, at about 
$c_0=7\rm \: mol\%$, while the substantial deviations from the linear dependence of 
$E_{ib}$ on concentration become apparent already at $c_0=0.1 \rm \: mol\%$
\cite{carl78electrical,takahashi82}. Note that doping below $1 \rm \: mol \%$ is usually 
considered as very dilute \cite{Scott2000ordering}. On the other hand, one should take
into account possible chemical restrictions on solubility of certain acceptors in the 
bulk of the grains \cite{Hans2008}. In any case, a concentration of about 
$1 \rm \: mol \%$ is high enough in the sense of the presented here space charge 
mechanism of screening of polarization which allows for good agreement with experiment 
\cite{carl78electrical,takahashi82,arlt88internal}.

The proposed two-dimensional model of charge migration can be extended to include 
further features and mechanisms which may influence aging. Note that the change of 
the $180^{\circ}$ domain walls to the $90^{\circ}$-walls in the sketch presented in 
Fig. \ref{grain} does not entail strong modification of the results and is reduced 
solely to the substitution of $\sigma_p = |\mathbf{P}_s|$ by 
$\sigma_p = |\mathbf{P}_s|/\sqrt{2}$. More important task is to take into account the 
possible change of the domain pattern during the charge migration. The metastable 
domain structure results from the compromise between the energy of the electrostatic
depolarization field and the energy of domain walls \cite{Kittel1946,Mitsui1953}. 
The relaxation of the electrostatic 
energy can trigger the change of the domain pattern including possible creation or 
disappearance of domain walls at the grain boundaries. This process, in turn, involves
the energy contribution of the mechanical stresses which has not been considered in this 
model as yet. One more idealization of the suggested model is an abrupt change of the 
polarization at the domain boundaries. Considering space distribution of the 
polarization within the Ginzburg-Landau approach reveals that chracteristic scale
at which the polarization gradually changes may become comparable with the domain width, 
especially when the temperature is not far from the temperature of the phase transition 
into the paraelectric state \cite{Lukyanchuk2008}. Including the ferroelastic 
interaction in the Ginzburg-Landau description may also substantially change the form of 
the domains providing appearance of the well known needle-shaped domains 
\cite{Salje1996}. These all additional features do not preclude, nevertheless, appearence
of strong local depolarization fields which present the crucial element of the actual
model of aging due to charged defects migration. Self-consistent analysis of the system 
evolution with many additional variables presents a very challenging task which may 
be addressed in the future. At the actual stage, only electrostatic arguments were 
observed so far which allow, however, comparison with the theory of defect dipole 
reorientation where only electrostatic contributions were included, too.

\section{\label{sec:acknowledgement}Acknowledgements}

Discussions with Nina Balke, Ruediger Eichel, Edwin Garcia, Xin Guo, Hans Kungl,
Igor Lukyanchuk, Doru Lupascu, Maxim Morozov, Ralf Mueller, Igor Pronin, Hermann Rauh, 
Jurgen Roedel, Don Smyth, Alexander Tagantsev, Reiner Waser and  Vadim Kirillovich 
Yarmarkin are gratefully acknowledged. This work was supported by the Deutsche 
Forschungsgemeinschaft through the Sonderforschungsbereich 595.  

\bibliography{apssamp}

\end{document}